\documentclass[aps,prr,reprint,groupedaddress,nofootinbib,longbibliography]{revtex4-1}

\usepackage{placeins}
\usepackage{amsmath,amsfonts,amssymb}
\usepackage{amsmath}
\usepackage{graphicx}
\usepackage[colorlinks=true, allcolors=blue]{hyperref}
\usepackage{siunitx}
\usepackage{changes}
\definechangesauthor[name={Tracy Northup}, color=purple]{TN}
\definechangesauthor[name={Dmitry Bykov}, color=orange]{DB}
\definechangesauthor[name={Lorenzo Dania}, color=magenta]{LD}
\newcommand{\stkout}[1]{\ifmmode\text{\sout{\ensuremath{#1}}}\else\sout{#1}\fi}
\setdeletedmarkup{\stkout{#1}}
\graphicspath{{Images/}}

\begin{document}


\title{Optical and electrical feedback cooling of a silica nanoparticle levitated in a Paul trap}


\author{Lorenzo Dania}
\email[]{lorenzo.dania@uibk.ac.at}
\affiliation{Institut f{\"u}r Experimentalphysik, Universit{\"a}t Innsbruck, Technikerstra\ss e 25, 6020 Innsbruck,
	Austria}
\author{Dmitry S. Bykov}
\email[]{dmitry.bykov@uibk.ac.at}
\affiliation{Institut f{\"u}r Experimentalphysik, Universit{\"a}t Innsbruck, Technikerstra\ss e 25, 6020 Innsbruck,
	Austria}
\author{Matthias Knoll}
\affiliation{Institut f{\"u}r Experimentalphysik, Universit{\"a}t Innsbruck, Technikerstra\ss e 25, 6020 Innsbruck,
	Austria}
\author{Pau Mestres}\affiliation{Institut f{\"u}r Experimentalphysik, Universit{\"a}t Innsbruck, Technikerstra\ss e 25, 6020 Innsbruck,
	Austria}
\author{Tracy E. Northup}
\affiliation{Institut f{\"u}r Experimentalphysik, Universit{\"a}t Innsbruck, Technikerstra\ss e 25, 6020 Innsbruck,
	Austria}



\date{\today}

\begin{abstract}
All three motional modes of a charged dielectric nanoparticle in a Paul trap are cooled simultaneously by direct feedback to temperatures of a few \si{\milli\kelvin}. We test two methods, one based on electrical forces and the other on optical forces; 
for both methods, we find similar cooling efficiencies. Cooling is characterized for both feedback forces as a function of feedback parameters, background pressure, and the particle's position.
\end{abstract}
\maketitle
\section{\label{sec:intro}introduction}
The Paul trap, a long-established tool for trapping atomic ions for quantum information science \cite{Leibfried2003}, has recently been harnessed as an optomechanics platform with which to trap charged dielectric micro- and nanoparticles and to cool their motion toward the quantum regime. In contrast to levitated optomechanics experiments based on optical tweezers traps, which require high laser intensities, in experiments based on Paul traps, light is used only to probe the motion of a trapped particle or to manipulate its internal and external degrees of freedom. This probing field can be significantly weaker than optical trapping fields, reducing both motional decoherence induced by random scattering of photons \cite{JainGieselerMoritzEtAl2016} and internal heating of the levitated dielectric particle \cite{millen2014nanoscale,Hebsestreit2018, Monteiro2020}, which in turn increases trapping stability and avoids irreversible damage to the sample. Moreover, Paul traps have deep potentials (\SI{\sim 1}{\kilo\electronvolt}~\cite{bykov2019}) and extended trapping regions (\SI{\sim 1}{\centi\meter^3}) with respect to optical tweezers (\SI{\sim 1}{\electronvolt}, \SI{1}{\micro\meter^3}~\cite{gieseler2012}), making Paul traps well-suited for the study of multi-particle or multi-species interactions \cite{ostendorf, Zhang2007,Willitsch2012}. These advantages establish Paul traps as a promising quantum optomechanical platform for testing spontaneous wave-function collapse models \cite{goldwater2016, vinante2019}, performing quantum free-fall experiments \cite{oriol2011_2}, generating non-classical states of motion \cite{oriol2010, chang2010}, and searching for evidence of the quantum nature of linearized gravity \cite{marshman2020}, provided that these experiments are performed in an ultra-high-vacuum (UHV) or cryogenic environment and that the particles are cooled to a low-temperature state.

Cooling of dielectric micro- and nanoparticles, pioneered in optical traps, has been achieved with several techniques, including direct feedback via cold damping \cite{ashkin1977feedback,li2011, iwasaki2019, conangla2019optimal, Felix_damping}, parametric feedback cooling \cite{gieseler2012}, and cavity-assisted resolved-sideband cooling \cite{kiesel2013}. The quantum realm has now been reached in optical-tweezers-based experiments, exploiting cavity cooling via coherent scattering \cite{uros2020} as well as feedback cooling via cold damping \cite{felix_as}. Beyond optical trapping experiments, cold damping has been used to cool microspheres in a magneto-gravitational trap \cite{Slezak2018} and has been applied to a wide range of physical systems, including micro-mechanical cantilevers \cite{pinard_2000, poggio2007feedback}, membranes \cite{rossi2018}, and single atomic ions \cite{Bushev2006}. In Paul traps, cooling of dielectric particles  has been demonstrated via parametric feedback on levitated graphene flakes \cite{nagornykh2015} and nitrogen-vacancy (NV) centers \cite{Conangla2018}, via spin cooling of librational modes of NV centers \cite{delord2020}, and via cavity cooling of silica nanospheres \cite{Millen2015}. To date, the best performance has been reported in a hybrid trap formed by an optical cavity overlapped with a Paul trap potential \cite{fonseca2016}.

In this work, we report the lowest temperature to date for a silica nanoparticle trapped in a Paul trap by using a cold damping technique. The particle is cooled simultaneously along all three trap axes using either an electrical or an optical feedback force, and we measure temperatures of the center-of-mass motion of a few millikelvin for both methods. With regards to optical cooling, we exploit both the gradient and the scattering forces of a single weak laser beam as a novel feedback actuator for cooling all modes of oscillation. Furthermore, we characterize both electrical and optical cooling performance in terms of all relevant parameters to which we have direct experimental access, namely, feedback gain and phase, background pressure, and particle position, which allows us to find an optimum regime to achieve the lowest temperature possible in our setup. Understanding the cooling process is important since it is an essential step for various levitated-optomechanics experiments~\cite{Millen_2020}.

The paper is organized as follows: In Section \ref{sec:apparatus}, we describe optical detection of the nanosphere's motion and the generation of electrical and optical feedback forces. In Section \ref{sec:cooling}, we present results on 3D feedback cooling, the performance of which is ultimately limited by the back-action of the particle position measurement. We also show how the single-axis cooling performance depends on the feedback gain, the feedback phase, and the background pressure. In Section \ref{sec:position}, exploiting the fact that the mechanisms for the Paul trap and for optical detection are decoupled, we show how cooling along the Paul trap axes depends on the position of the particle with respect to both the detection laser beam and the fields of feedback forces. We summarize our findings in Section \ref{sec:conclusion}.

\section{\label{sec:apparatus}Description of the experimental apparatus}
\begin{figure*}[ht]
	\centering
	\includegraphics[width=1\linewidth]{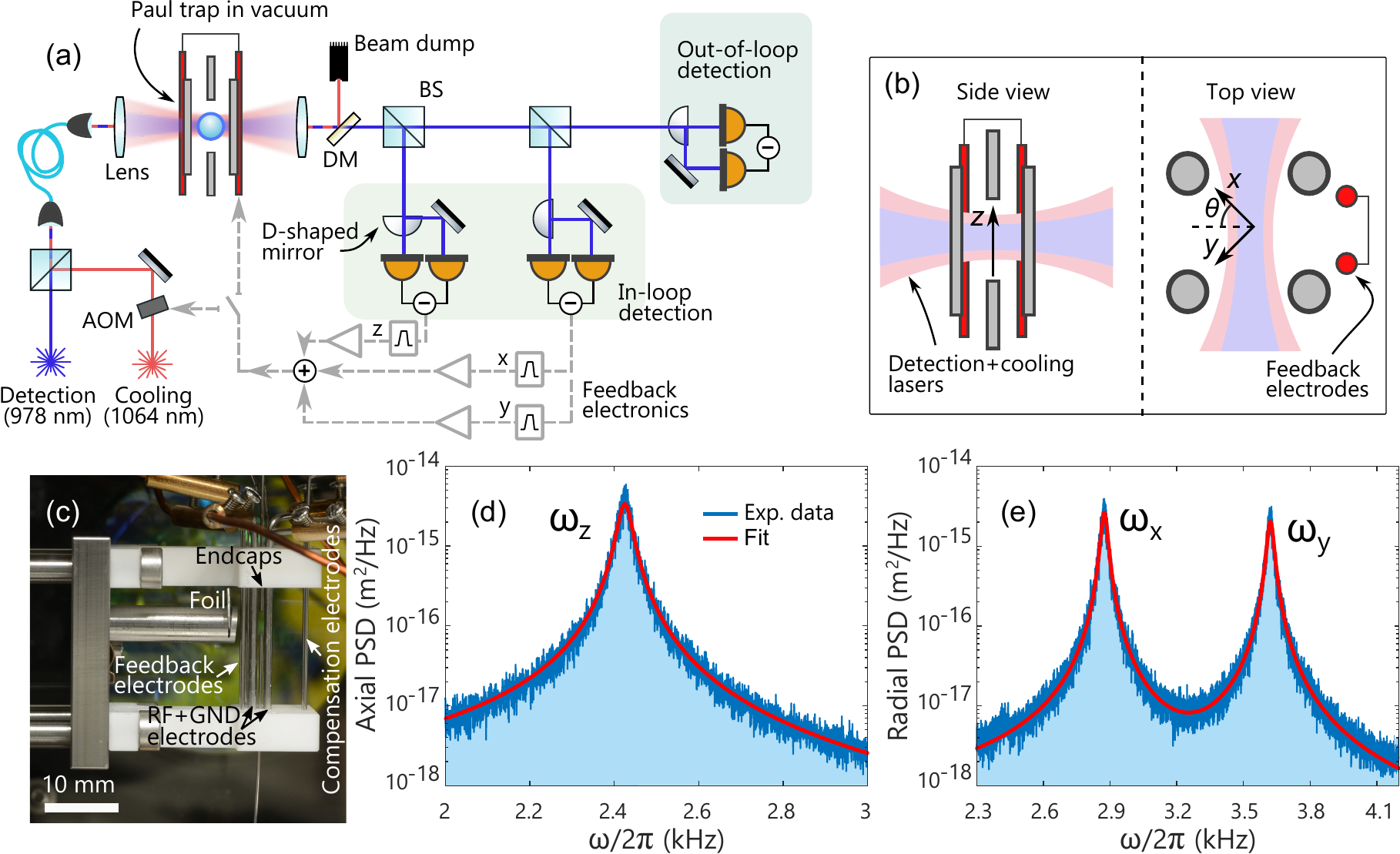}
	\caption{(a) Schematic view of the detection and cooling setup for nanospheres: a \SI{978}{\nano\meter} detection laser is focused onto a trapped nanosphere. Forward-scattered light is collected by a lens and equally split in three branches by a pair of beam splitters (BS). Each beam is sent to a balanced photodetector which measures the particle's vertical or horizontal motion (with respect to the optical table), depending on the orientation of the D-shaped mirror in front of each detector. The signals from the in-loop detectors are processed by feedback electronics, which consist of three FPGA cards, and then sent to the feedback force actuators. For optical cooling, the feedback signal is sent to an acousto-optic modulator (AOM), which modulates the intensity of a \SI{1064}{\nano\meter} cooling laser. The cooling laser follows the same optical path used for the detection laser up to the levitated particle, after which it is decoupled from the detection path with a dichroic mirror (DM). For electrical cooling, the feedback signal is sent to the feedback electrodes of the Paul trap. A switch allows us to select which cooling method is applied. (b) Orientation of the Paul trap's axes with respect to the laser beams. (c) Image of the Paul trap's electrodes and its mounting system in the vacuum chamber. The horizontal tube is part of the particle's loading apparatus~\cite{bykov2019}. (d) Example of the measured single-sided power spectral density (PSD) of the particle's motion obtained from the axial detector. The red line is a fit to a Lorentzian function peaked at the axial oscillation frequency $\omega_z$. (e) Example of the measured PSD of the particle's radial motion. The red line is a fit to a double-peaked Lorentzian with the radial oscillation frequencies $\omega_x$ and $\omega_y$. More details about the fits can be found in Appendix~\ref{ap:detection}.}
	\label{fig:setup}
\end{figure*}

$\text{Fig.}\, \ref{fig:setup}${(a)} shows a sketch of the experimental setup. We trap single silica nanospheres, \SI{300}{\nano\meter} in diameter, in a four-rod linear Paul trap. The mass of the particles is on the order of \SI{e-17}{\kilo\gram}; our mass measurement procedure for individual particles is described in  Appendix~\ref{ap:jumps}. A laser-induced acoustic desorption (LIAD) technique~\cite{AsenbaumKuhnNimmrichterEtAl2013,Millen2016}, applied in a time-controlled manner, is used to charge particles and to load them into the trap in high vacuum~\cite{bykov2019}.

The four cylindrical rods of the Paul trap are arranged in a square configuration and provide confinement in the plane perpendicular to the trap axis; we refer to this as the radial or $xy$ plane.  The diagonal distance between electrodes is $2r_0 = \SI{1.8}{\milli\meter}$. Two additional DC-biased electrodes, referred to as endcaps, provide confinement along the ion-trap axis, $z$. The distance between the endcaps is $2z_0 = \SI{3.4}{\milli\meter}$. The orientation of the trap axes is depicted in $\text{Fig.}\, \ref{fig:setup}${(b)}, and an image of the trap mounted in the vacuum chamber is shown in $\text{Fig.}\, \ref{fig:setup}${(c)}. For the results presented here, the trap is driven at frequencies between \SI{10}{\kilo\hertz} and \SI{15}{\kilo\hertz}, with amplitudes ranging from \SI{400}{\volt} to \SI{800}{\volt}.  Typical endcap voltages lie between +\SI{150}{\volt} and +\SI{270}{\volt}. With these parameters, we obtain motional frequencies of a few kilohertz along all three trap axes.

We measure a particle's motion using homodyne detection~\cite{Gittes1998}: \SI{1}{\milli\watt} of \SI{987}{\nano\meter} laser power is focused at the particle's position using an objective mounted outside of the vacuum chamber on a three-axis translation stage. The forward transmitted light and the scattered light from the particle are then collected by a 0.3 NA objective and divided in three paths by two beam splitters. Each of the three beams is split by a D-shaped mirror, and the two components of each beam are then steered to opposite ports of a balanced photodiode, such that the resultant photocurrent is proportional to the particle's displacement from the trap center along a given axis.  Two of the balanced photodiodes are used for axial and radial detection as part of the cooling feedback loop, while the third is used as an out-of-loop detector to calibrate and measure the particle temperature. Typical power spectral densities (PSDs) of the particle motion measured along the axial and radial trap axes at \SI{3.8e-2}{\milli\bar} are shown in $\text{Figs.}\, \ref{fig:setup}${(d) and (e)}, respectively. The resonant frequencies in this case are $(\omega_x, \omega_y, \omega_z)/(2\pi)=( 2.87, 3.63, 2.42)\,\si{\kilo\hertz}$.

The radial in-loop signal contains information about particle motion along both the $x$ and $y$ axes.  In order to implement individual feedback control along each axis, the radial signal is split in two, and tunable bandpass filters isolate the $x$ motion in one path and the $y$ motion in the other.  A filter is also implemented on the axial ($z$) signal.  Each of the three signals is then amplified and phase shifted to obtain a feedback signal proportional to the particle's velocity along the respective axis.  The feedback electronics are implemented by field-programmable gate arrays (FPGAs)~\cite{neuhaus2017pyrpl}.
The three feedback signals are then combined and sent to the feedback force actuators.

Feedback forces on the particle are applied either electrically or optically. Electrical feedback is provided via a pair of electrodes oriented nearly parallel to the Paul trap axis ($\text{Fig.}\, \ref{fig:setup}(\text{b})$). When electrical cooling is on, the field generated by the electrodes is proportional to the amplitude of the feedback signal set with the {FPGAs}. At maximum gain, the voltage at the feedback electrodes does not exceed \SI{10}{\volt}. The electric field generated by the feedback electrodes has a projection along all axes of motion, enabling cooling in three dimensions. Optical feedback is achieved by modulating the intensity of a \SI{1064}{\nano\meter} laser beam with an acousto-optic modulator (AOM). This cooling beam is coupled to the particle via the same objective used for the detection laser. As depicted in $\text{Fig.}\, \ref{fig:setup}(\text{b})$, the cooling beam intersects both the $x$ and $y$ axes at \SI{45}{\degree} and the $z$ axis at \SI{90}{\degree}, so that the radiation pressure force pushes the particle along both the $x$ and $y$ axes, while the gradient force pulls it towards the beam focus along the $z$ axis. When cooling is off, the \SI{1064}{\nano\meter} power at the particle is \SI{4}{\milli\watt}, and the intensity is \SI{2e5}{\watt\per\cm^2}, which can be compared to typical values of \SI{5e7}{\watt\per\cm^2} for particles optically trapped in tweezers~\cite{gieseler2012}.  When optical cooling is on, the cooling beam power is modulated with an amplitude proportional to the feedback gain, up to \SI{100}{\percent} of the unmodulated beam power.

We emphasize that in the presence of both the cooling and detection beams, the particle's motional frequencies are modified by less than \SI{1}{\percent} due to the gradient forces of the beams, ensuring that confinement is due to the Paul trap potential and that the optical forces act as small perturbations.

\section{\label{sec:cooling}Optical and electrical feedback cooling}

\subsection{\label{sec:cooling3axes}Cooling along three axes}

There are three features of the experimental setup that allow us to perform cooling in three dimensions.  First, both electrical and optical feedback forces have non-zero projections along all trap axes. Second, we can individually tune the phase and the gain of the feedback signal for each center-of-mass mode of oscillation of the nanosphere. Third, each mode of oscillation can be accessed individually, thanks to the resonant interaction between the particle motion and the feedback force.

\begin{figure}[ht]
	\centering
	\includegraphics[width=1\linewidth]{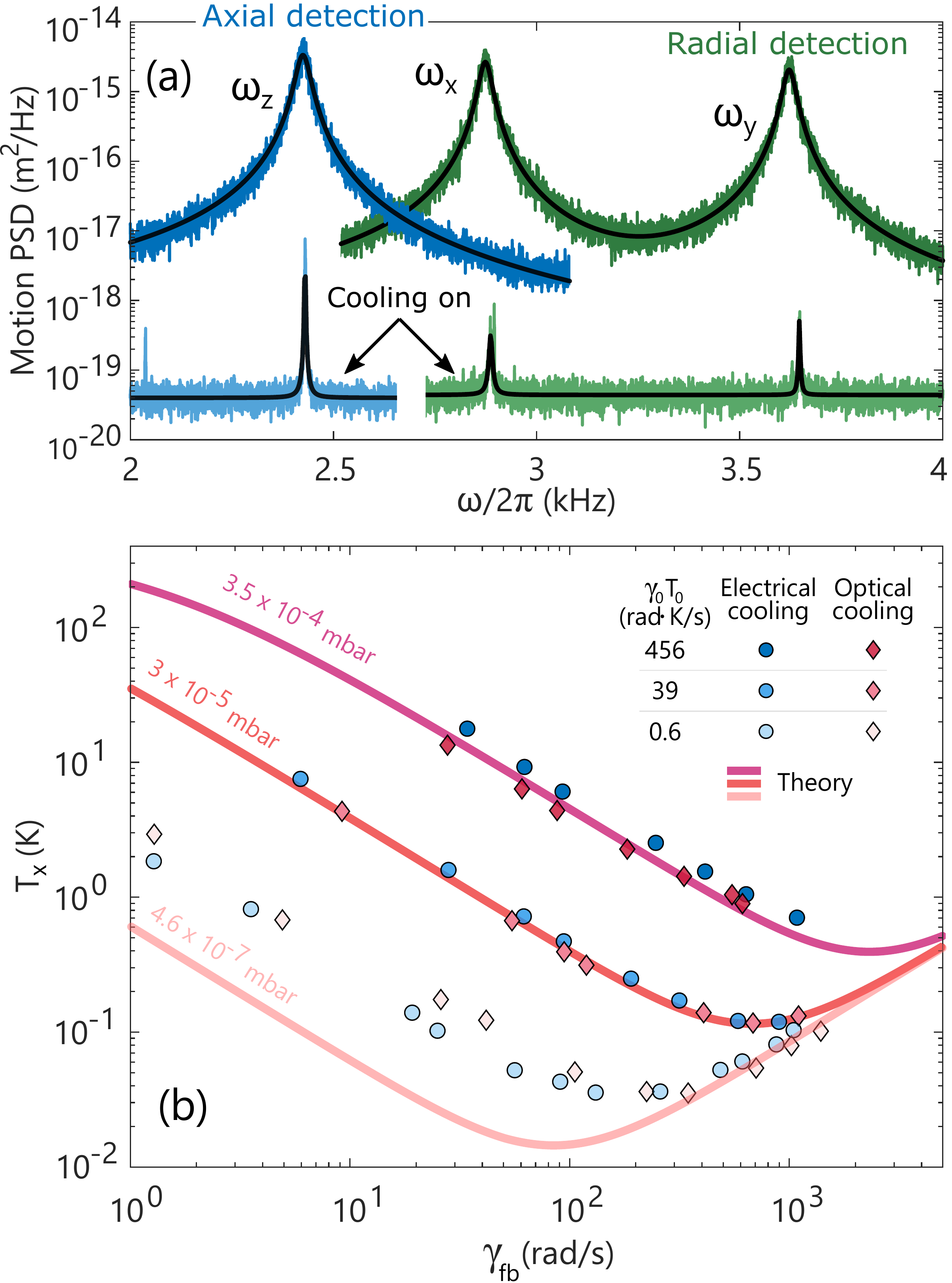}
	\caption{(a) 3D cooling obtained with optical feedback. The figure shows axial (blue) and radial (green) PSDs of the particle's motion taken at a pressure of \SI{3.8e-2}{\milli\bar} with feedback off (dark) and at \SI{2.5e-7}{\milli\bar} with optical cooling on (light). Black lines are fits to Lorentzian functions. Data taken at low pressure correspond to temperatures of $T_z=\SI{31\pm4}{\milli\kelvin}$, $T_x=\SI{6\pm1}{\milli\kelvin}$ and $T_y=\SI{6\pm1}{\milli\kelvin}$. The additional sharp peak near $\SI{2}{\kilo\hertz}$ in the cooled spectrum are due to feedback electronics and laser intensity noise. (b) Steady-state temperature of the motion along the $x$\protect\nobreakdash-axis ($T_x$) as a function of feedback gain $\gamma_{\text{fb}}$ under electrical (blue circles) and optical (red diamonds) cooling, taken at different pressures. The error bars are smaller than the data points. Lines are plots of $\text{Eq.}\, (\ref{eq:TvsG})$ without any free parameters.
	The discrepancy between the lowest pressure data and theory suggests that the buffer gas in this experimental situation is no longer the main source of heating. For each data point, the phase shift between the detection signal and the applied feedback force was set to an optimum value as described in Sec.~\ref{sec:cooling:phase}.}
	\label{fig:tvs}
\end{figure}

$\text{Fig.}\, \ref{fig:tvs}(\text{a})$ shows typical PSDs of a nanosphere's center-of-mass motion with and without 3D optical feedback cooling.  The broader set of PSDs, shown in dark colors, are taken at a pressure of \SI{3.8e-2}{\milli\bar} with feedback off in order to calibrate the detection system. At this pressure, the particle thermalizes to an equilibrium temperature $T_0$, which in the case of a simple harmonic oscillator would be equal to the temperature of the buffer gas. However, because particles in the Paul trap are subject to a time-dependent potential, $T_0$ does not necessarily match the background-gas temperature~\cite{chen2014neutral}. For the measurements presented in this paper, $T_0$ in the $xy$ plane is $1.3$ times higher than the temperature of the gas, which is at $\SI{300}{\kelvin}$. A more detailed discussion on the equilibrium temperature in the absence of feedback cooling can be found in  Appendix \ref{ap:calib_temp}. The narrower set of PSDs, shown in light colors, are taken at \SI{2.5e-7}{\milli\bar} with feedback on. From fits to the axial and radial PSDs, we extrapolate temperatures of $(T_x,T_y,T_z)=(8(1),8(1),31(4))$~\si{\milli\kelvin}. With electrical feedback, we carry out a similar analysis and extrapolate temperatures of $(T_x,T_y,T_z)=(9(1),8(1),7(1))$~\si{\milli\kelvin}.
More details on the fits can be found in Appendix \ref{ap:detection}.
The larger value of $T_z$ obtained under optical cooling is due to a lower feedback gain used in that particular experimental realization, which we found in retrospect was not optimized.
These temperatures are three orders of magnitude lower than those previously reported for levitated nanoparticle cooling experiments performed in a Paul trap potential~\cite{nagornykh2015,Conangla2018,delord2020} and similar to temperatures achieved along one axis in hybrid electro-optical traps~\cite{fonseca2016}.

\subsection{Single-axis cooling performance}

\subsubsection{Dependence on the feedback gain}

A particle's motion along one axis is decoupled from its motion along orthogonal axes, allowing us to focus on single-axis performance to find optimal cooling parameters for each axis.  Here, we study single-axis cooling under both optical and electrical feedback for a range of parameters, including pressure, feedback gain, and feedback phase.

In $\text{Fig.}\, \ref{fig:tvs}(\text{b})$, we plot the steady-state temperature $T_x$ reached using either electrical or optical feedback as a function of the cooling rate $\gamma_{\text{fb}}$ and for three values of the background gas pressure. The cooling rate is defined as the ratio of the feedback force $F_{\text{fb}}$ to the particle's momentum:
\begin{equation}
\gamma_{\text{fb}} = \frac{F_{\text{fb}}}{mv(t)},
\label{eq:gammafb}
\end{equation}
where $m$ is the nanosphere's mass and $v(t)$ is the nanosphere's velocity. We measure $\gamma_{\text{fb}}$ as the linewidth of the out-of-loop PSD of the particle's motion along the $x$ axis; this linewidth is proportional to the feedback gain set with the FPGA card.
We find that the temperatures measured under optical cooling do not significantly differ from those measured under electrical cooling for the same pressure and feedback parameters. This result suggests that the cooling laser does not introduce significant noise, and in particular, that neither radiation-pressure shot noise nor laser-intensity fluctuations plays a significant role.

We compare our single-axis experimental results to the prediction of the cold damping model of Ref.~\cite{Felix_damping}, which states that
\begin{equation}
T_x=\frac{\gamma_0T_0}{\gamma_0+\gamma_{\text{fb}}}+\frac{\pi m\omega_x^2}{2k_B}\frac{\gamma^2_{\text{fb}}}{\gamma_0+\gamma_{\text{fb}}}S_{\delta x_{\text{il}}},
\label{eq:TvsG}
\end{equation}
where $\gamma_0$ is the pressure-dependent natural linewidth of the oscillator, $k_B$ is the Boltzmann constant, and $S_{\delta x_{\text{il}}}$ is the single-sided PSD of the position measurement noise as measured by the in-loop detector. Note that this theory assumes that collisions with the background gas are the reheating mechanism by which the particle thermalizes at $T_0$ in the absence of feedback. Equation~$(\ref{eq:TvsG})$ is plotted with solid lines in $\text{Fig.}\, \ref{fig:tvs}(\text{b})$ for the three experimental pressure values.

All parameters that enter $\text{Eq.}\, (\ref{eq:TvsG})$ are independently estimated in our experiment: The mass $m = \SI{1.8(2)e-17}{\kilo\gram}$ is extracted from discrete jumps of the nanosphere's charge, a method that we describe in Appendix \ref{ap:jumps}.  The PSD $S_{\delta x_{\text{il}}}$ is directly measured by the in-loop detector, and the motional frequency $\omega_x$ is determined from the fit of the PSD measured at the out-of-loop detector. The linewidth $\gamma_0$ is measured at the calibration pressure of \SI{3.8e-2}{\milli\bar} with feedback cooling off and then deduced for all other pressures $P$ under the assumption that in our system, the relation $\gamma_0\propto~P$~\cite{Beresnev1990,JainGieselerMoritzEtAl2016,bykov2019} holds.

At pressures of \SI{3.5e-4}{\milli\bar} and \SI{3.5e-5}{\milli\bar}, the data and theory curves are in excellent agreement. At all three pressures shown in $\text{Fig.}\, \ref{fig:tvs}(\text{b})$, the relation $\gamma_0 \ll \gamma_{\text{fb}}$ holds; in this limit, the first term on the right-hand side of $\text{Eq.}\, (\ref{eq:TvsG})$ predicts a decrease of the temperature $T_x$ inversely proportional to $\gamma_{\text{fb}}$. This term acts to cool the particle below the background temperature  $T_0$, and for low feedback gain, we observe the inverse dependence on $\gamma_{\text{fb}}$ for both sets of experimental data. For high feedback gain, the second term on the right-hand side of $\text{Eq.}\, (\ref{eq:TvsG})$ becomes dominant.  In the limit $\gamma_0 \ll \gamma_{\text{fb}}$, this term predicts a temperature increase linearly proportional to $\gamma_{\text{fb}}$ and thus the appearance of a minimum of $T_x$. The temperature increases because the feedback amplifies not only the position measurement but also the unwanted noise of the measurement, introducing heating into the system. For the data taken at \SI{3.5e-5}{\milli\bar}, we observe a minimum of $T_x$, in accordance with the theoretical curve.  We also observe noise squashing at the in-loop detection system, a known signature of the limit of the cold damping technique~\cite{Bushev2006,poggio2007feedback,rossi2018,Felix_damping,conangla2019optimal,pinard_2000}.

At \SI{4.6e-7}{\milli\bar}, in contrast, the temperatures obtained for low values of $\gamma_{\text{fb}}$ are higher than expected from a theory based solely on a gas reheating mechanism. This discrepancy with $\text{Eq.}\, (\ref{eq:TvsG})$  is not present on the detection-limited side of the curve, for high values of $\gamma_{\text{fb}}$, suggesting that the extrapolated $\gamma_0T_0$ coefficient may have been underestimated.
Moreover, we have found in subsequent measurements that the temperature at this pressure for low values of $\gamma_{\text{fb}}$ is sensitive to the position of the nanoparticle in the trap.
One possibility is that $\gamma_0$ at these pressures is determined not only by background-gas collisions but also by a position-dependent heating mechanism, such as excess particle micromotion \cite{berkeland98} or radiofrequency noise of the Paul trap~\cite{Blatt_RFheating}. The micromotion analysis presented in Appendix~\ref{ap:micromotion} shows that for our experimental parameters, the excess micromotion does not significantly influence the effective temperature of the secular motion. Moreover, in our experiments, the nanoparticle position was optimized to minimize the micromotion. Therefore, we suspect the radiofrequency noise as a source of the discrepancy in $\text{Fig.}\, \ref{fig:tvs}(\text{b})$, but this effect needs further investigation.

\subsubsection{\label{sec:cooling:1axis:min}Minimum temperature}

The results in Sec.~\ref{sec:cooling3axes} correspond to mean phonon occupation numbers $\bar{n}_i=k_BT_i/\hbar\omega_i$ between \SI{4.6(8)e4}{} and \SI{3.4(4)e5}{}, where $i\in \{x,y,z\}$ and $\hbar$ is the reduced Planck constant. Given the recent cooling of nanoparticles to the quantum regime \cite{uros2020,felix_as}, it is important to understand what limits further cooling in this setup.

Minimizing $T_x$ in $\text{Eq.}\, (\ref{eq:TvsG})$ with respect to $\gamma_{\text{fb}}$, we see that the minimum temperature scales as $T^{\text{min}}_x\propto\sqrt{\gamma_0T_0S_{\delta x_{\text{il}}}}$.
Replacing lossy optics in the detection path and substituting a single quadrant photodetector for the three balanced detectors would improve $S_{\delta x_{\text{il}}}$ by a factor of $10^2$ and $\bar{n}_i$ by a factor of 10. Moreover, we calculate that switching from detection of forward-scattered light to a scheme based on self-interference detection \cite{Bushev2013} would provide another two orders of magnitude improvement in $S_{\delta x_{\text{il}}}$ (manuscript in preparation).  To reduce the occupation number even further at room temperature, one could reduce the background pressure $\gamma_0$ or increase the trap frequency $\omega_i$.  In the latter case, increasing the number of charges on the nanoparticle or decreasing the particle's mass will allow us to trap stably at higher frequencies. Finally, one could introduce cryogenic buffer gas to reduce $T_0$ and, therefore, the minimum temperature.

\subsubsection{\label{sec:cooling:phase}Dependence on the feedback phase}
\begin{figure}[ht]
	\centering
	\includegraphics[width=1\linewidth]{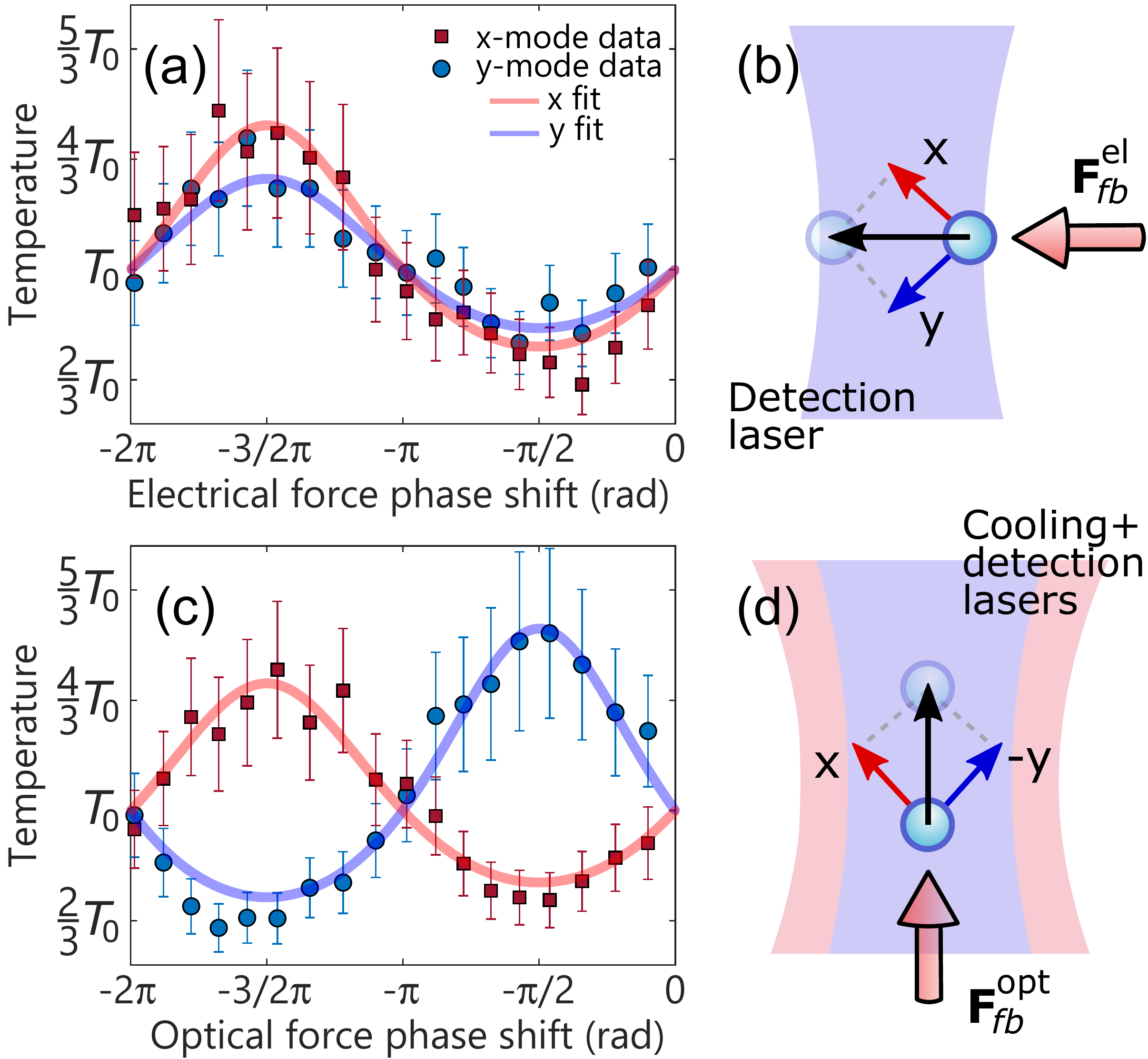}
	\caption{(a) Temperatures $T_x$ and $T_y$  of the $x$ (red) and $y$ (blue) modes of the particle's oscillation under electrical feedback cooling as a function of the feedback force's phase shift $\phi$ with respect to the particle's motion. (b) The electrical force $\textbf{F}^{\text{el}}_{\text{fb}}$ due to the feedback electrodes points along $\hat{x}+\hat{y}$. (c) $T_x$ (red) and $T_y$ (blue) under optical cooling as a function of $\phi$. (d) The force $\textbf{F}^{\text{opt}}_{\text{fb}}$ due to radiation pressure from the cooling laser has a projection along $\hat{x}-\hat{y}$. As a result, the optimal phase for optical feedback along the $y$\protect\nobreakdash-axis is shifted by $\pi$ with respect to the optimal phase for electrical feedback.}
	\label{fig:radial_phase}
\end{figure}
We apply a feedback force that is proportional to the phase-shifted position signal of the particle.  The nanosphere cooling performance depends not only on the amplitude of the force but also on the tunable phase shift~$\phi$, which is controlled by the output of the FPGA. Assuming the particle's position can be written as $x(t)=A\cos(\omega_xt)$, with $A$ the amplitude of motion, and using Eq.~\ref{eq:gammafb}, the feedback force can be written as
\begin{equation}
\begin{split}
	F_{\text{fb}}&= m\gamma_{\text{fb}}{\omega_x}A\cos(\omega_xt+\phi)\\
    &=m\omega^2_{\text{fb}} \cos(\phi) x(t) + m\gamma_{\text{fb}} \sin(\phi) v(t),
	\label{eq:F_vs_phase}
	\end{split}
\end{equation}
where we define $\omega^2_{\text{fb}} = \gamma_{\text{fb}}{\omega_x}$, and we have decomposed the force into the $x(t)$ and $v(t)=\dot{x}(t)$ quadratures of motion. The same relationship holds for the $y$ and $z$ motion. From $\text{Eq.}\, (\ref{eq:F_vs_phase})$, it can be seen that $F_{\text{fb}}$ becomes a purely cooling force for $\phi^{\text{cool}}=-\frac{\pi}{2}$ and a purely heating force for $\phi^{\text{heat}}=-\frac{3}{2}\pi$. In all other cases, $F_{\text{fb}}$ acquires a term proportional to the particle's position. The proportionality coefficient $\omega^2_{\text{fb}}$ expresses the shift of the nanosphere's motional frequency along a given axis. For all measurements shown in this paper, these frequency shifts are small when compared to the bare nanosphere's motional frequencies, and so we will neglect this effect from here on and will consider only the term of $\text{Eq.}\, (\ref{eq:F_vs_phase})$ that is proportional to the velocity of the particle. The radial temperature $T_x$ as a function of $\phi$ is given by:
\begin{equation}
T_x(\phi)=\frac{\gamma_0T_0}{\gamma_0-\gamma_{\text{fb}}\sin(\phi)};
\label{eq:T_vs_phase}
\end{equation}
the derivation is outlined in Appendix \ref{ap:phase}. An analogous formula holds for the $y$ mode.

We measured $T_x(\phi)$ and $T_y(\phi)$ at \SI{4e-3}{\milli\bar} for both electrical and optical feedback. For both kinds of feedback forces, the phase is measured relative to the particle's position signal. The feedback gain was kept low to avoid losing the particle when the phase corresponds to particle heating. Data for the case of electrical feedback are shown in $\text{Fig.}\, \ref{fig:radial_phase}(\text{a})$. The solid lines in the plot are fit to $\text{Eq.}\, (\ref{eq:T_vs_phase})$, with $\gamma_{\text{fb}}$ as the only free parameter. For both radial modes, we observe cooling for $-\pi<\phi<0$ and heating otherwise. It is important to note that the electrical feedback force points along $\hat{x}+\hat{y}$ ($\text{Fig.}\, \ref{fig:radial_phase}(\text{b})$).
The data for $T_x$ and $T_y$ as a function of the optical force's phase are shown in $\text{Fig.}\, \ref{fig:radial_phase}(\text{c})$. In this case, the optical feedback force points along $\hat{x}-\hat{y}$ ($\text{Fig.}\, \ref{fig:radial_phase}(\text{d})$), due to the fact that the radiation pressure force of the cooling beam points toward the beam's propagation axis. As a consequence, $F^{\text{opt}}_x\propto x$ and $F^{\text{opt}}_y\propto -y$, so that for optical cooling, $\phi^{\text{cool}}_x=\phi^{\text{cool}}_y+\pi=-\frac{\pi}{2}$. This phase flip between electrical and optical cooling is also included in the fit to the $y$-mode data in $\text{Fig.}\, \ref{fig:radial_phase}(\text{c})$.

For minimizing crosstalk between the optical feedback forces on the two radial modes, it is crucial to filter the detection signal at each radial mode frequency with a bandwidth $\delta_{BW}$ small enough to fulfill the condition $\delta_{BW}<(\omega_y-\omega_x)/2$.

\section{\label{sec:position}Position-dependent detection and cooling efficiency}
In the previous section, we showed how the minimum temperature achieved with feedback cooling is ultimately limited by the back-action of the position measurement, which injects noise into the system. One strategy for increasing the signal-to-noise ratio for the position measurement, and therefore for obtaining better cooling performance, is to optimize the alignment of the detection laser to the particle's position in the Paul trap. This problem is unique to experiments in which the trapping mechanism (here, a Paul trap) and the detection mechanism are decoupled; in contrast, in optical tweezers experiments, the same optical beam traps a particle and detects its motion, so that alignment is guaranteed.
As discussed below, we have found that when the particle is displaced from the detection laser focus, the feedback cooling performance is reduced, and in some cases, the nanosphere is heated rather than cooled.

\subsection{Optimization of position detection}{\label{sec:position_optimization}
In order to maximize the particle signal, we have studied the response of the photodiodes to displacements of the nanosphere. We align the detection laser to the particle in two steps: First, we roughly align the laser to the nanosphere by moving the focusing lenses with a 3D translation stage while maximizing the scattered light of the particle imaged on a CMOS camera. Second, we finely adjust the particle's position by applying DC fields on two sets of Paul-trap electrodes: We displace the nanosphere along the $z$ axis by unbalancing the voltage on the endcap electrodes, and we displace the nanosphere in the radial plane via a set of compensation electrodes. 
The photodiodes are sensitive to particle movements in the laser's focal plane, so we focus on displacements in the $(\hat{u},\hat{z})$ plane, where we define $\hat{u} \equiv (\hat{x}+\hat{y})/\sqrt{2}$.

The signal $V_r(u)$ from the radial photodiode relative to the instantaneous particle position $u$ is~\cite{Gittes1998}
\begin{equation}
V_{r}(u)\propto\Big(\frac{u-u_0}{w_0}-\frac{2}{3}\Big(\frac{u-u_0}{w_0}\Big)^3\;\Big)e^{-((u-u_0)/w_0)^2},
\label{eq:gittes}
\end{equation}
where $w_0$ is the electric-field waist (1/$e$ radius) of the laser and $u_0$ is the distance between the focus and the geometrical center of the Paul trap. An analogous relationship exists for the axial photodiode signal $V_z(z)$. The functions $V_r(u)$ and $V_z(z)$ give the value of the particle's motional signal in volts in response to particle displacements $u$ and $z$, respectively, in micrometers. For particle oscillations of amplitude $\delta u$ smaller than the waist $w_0$, the response function can be linearized around the particle's center of oscillation $u_{\text{eq}}$ as

\begin{equation}
	\delta V_r(u_{\text{eq}})=\frac{\partial V_r}{\partial u}\Big|_{u_{\text{eq}}}\delta u.
	\label{eq:linearized_gittes}
\end{equation}
$\text{Equation}\, (\ref{eq:linearized_gittes})$ states that to first order, the slope of the detector's response function is the conversion factor between the real amplitude of motion $\delta u$ and the measured amplitude $\delta V_r$. Moreover, this conversion factor depends on the equilibrium position $u_{\text{eq}}$ of the particle in the trap, meaning that the measured amplitude of motion depends on the particle's displacement in the Paul trap. We refer to $c_r(u_{\text{eq}})=\delta V_r/\delta u$ as the radial position-dependent conversion factor; an analogous relation exists for the axial conversion factor $c_z(z_{\text{eq}})=\delta V_z/\delta z$. In order to maximize both signals, $\delta V_r(u_{\text{eq}})$ and  $\delta V_z(z_{eq})$, the particle has to be displaced to a point at which both the axial and radial conversion factors have a maximum, which is fulfilled if $u_{\text{eq}}$ and $z_{eq}$ coincide with the laser focus.

We characterize our system by determining both $V_r(u)$ and $V_z(z)$. To obtain  $V_r(u)$, we first displace the particle over a range of equilibrium positions $u_{\text{eq}}$ and measure the conversion factor $c_r$ for each value of $u_{\text{eq}}$. The absolute value of each conversion factor $|c_r|$ is determined from a fit to the PSD of the particle's motion~\cite{hebestreit2018}, and the sign is determined by measuring the phase response to a harmonic driving force (Appendix~\ref{ap:conversion_factor}).  Next, using $\text{Eq.}\, (\ref{eq:linearized_gittes})$, we extract the response function $V_r(u)=\int_{0}^{u}c_r(u_{\text{eq}})du$. An analogous procedure is used to obtain $V_z(z)$.

$\text{Figures}\, \ref{fig:detection}\text{(a) and (b)}$ show the experimental results for $V_r(u)$ and $V_z(z)$, respectively. The data are fitted to $\text{Eq.}\, (\ref{eq:gittes})$ with $w_0$, the laser focus position, and a global prefactor as free parameters. The values for the displacements $u$ and $z$ in microns are extrapolated from simulations of the electric field generated by the compensation electrodes, where we have independently determined the charge state of the nanosphere (Appendix~\ref{ap:jumps}). We see from $\text{Figs.}\, \ref{fig:detection}\text{(a) and (b)}$ that the response functions are linear around the focus for a range of particle amplitudes up to $w_0\approx\SI{3}{\micro\meter}$. For comparison, when feedback cooling is off, the mean amplitude of oscillation for the $z$ mode is $\delta z=\sqrt{k_BT/(m\omega_z^2)}=\SI{0.9(1)}{\micro\meter}$, while for the radial modes, we have $\delta u_x=\SI{0.74(6)}{\micro\meter}$ and $\delta u_y=\SI{0.51(6)}{\micro\meter}$, where we have taken the projection of the $x$ and $y$ modes along the detection direction $\hat{u}$. The fact that $\delta z,\delta u_x,\delta u_y < w_0$ ensures that when the particle is placed in the detection beam focus, its motion along all trap axes is linearly mapped to the detection signal.

For amplitudes larger than $w_0$, the response functions become nonlinear.  As a result, the particle signal becomes smaller, and higher harmonics appear on the signal's spectrum. Note that for $|u|,|z|\gg w_0$, the response functions tend to zero, given that the position signal vanishes in the limit that the particle is pushed outside the detection beam. The signs of the derivatives of $V_r$ and $V_z$ flip accordingly. In $\text{Figures}\, \ref{fig:detection}\text{(a) and (b)}$, we observe, in agreement with $\text{Eq.}\, (\ref{eq:gittes})$, that $V_r$ and $V_z$ each have two additional linear zones, for which the sign of the slope is opposite to the sign of the slope around the focus. Through $\text{Eq.}\, (\ref{eq:linearized_gittes})$, the sign flip of the response function is mapped to a sign change in the conversion factor $c_r$: the feedback force flips sign as the particle is moved across the border between two linear zones of the response function.  An equivalent statement holds for $c_z$.

\subsection{\label{sec:position_vs_T}Position-dependent cooling performance}
\begin{figure}[ht]
	\centering
	\includegraphics[width=1\linewidth]{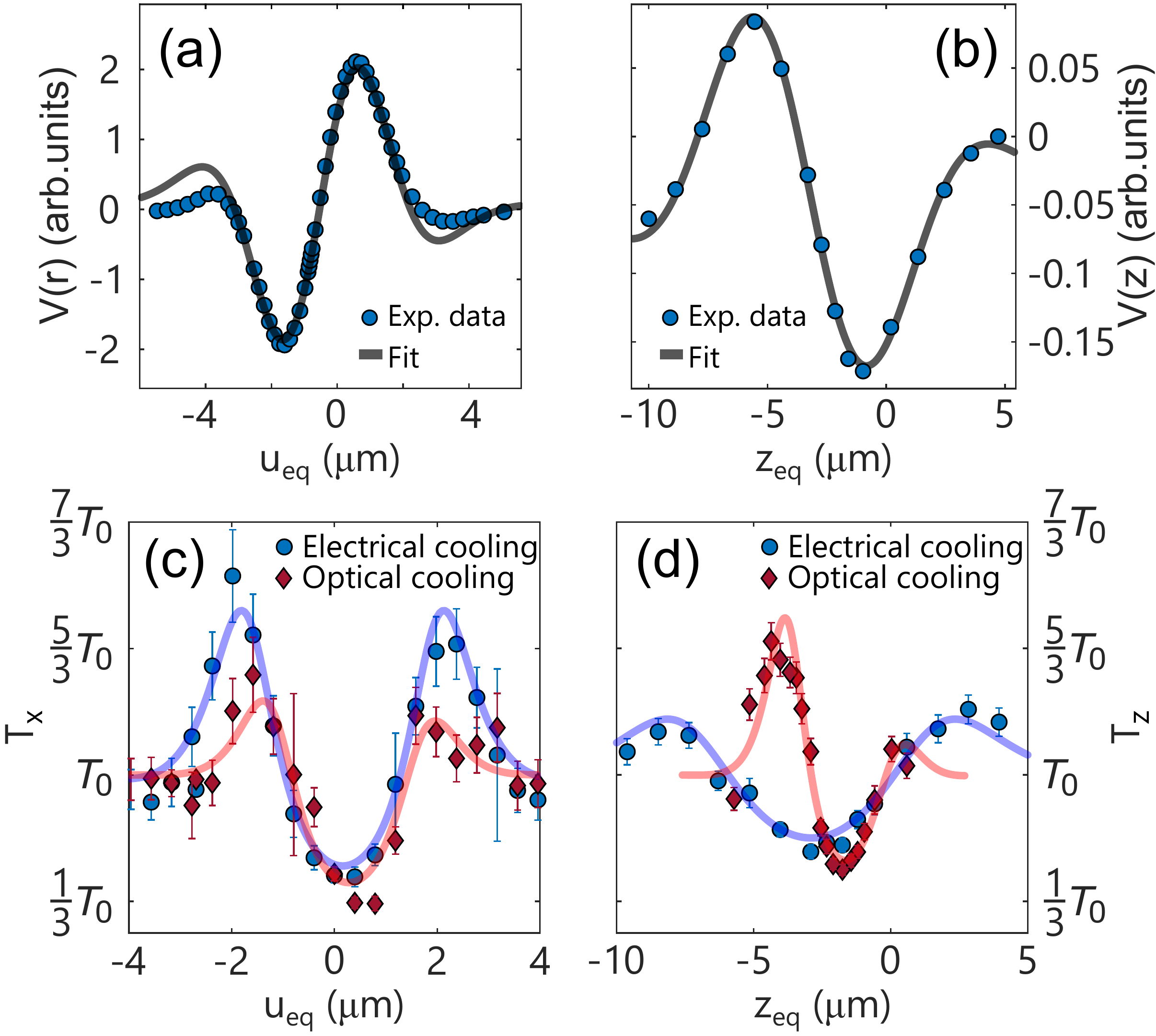}
	\caption{(a) Measured radial detector response $V$ as a function of the particle's displacement along the $u$ axis. The line is a fit to $\text{Eq.}\, (\ref{eq:gittes})$. (b) Axial detector response $V(z)$ as a function of the particle's axial displacement $z$.  The line is a fit to $\text{Eq.}\, (\ref{eq:gittes})$ with $u$ and $u_0$ replaced by $z$ and $z_0$.  The position $(u,z)=(0,0)$ corresponds to the geometrical center of the trap. Real displacements are calculated from electrostatic simulations and from the known charge state of the particle. (c)  Steady-state temperature $T_x(u_{\text{eq}})$ obtained under electrical and optical cooling (blue circles and red diamonds, respectively). Lines are fits to $\text{Eq.}\, (\ref{eq:T_vs_position})$. (d) Steady-state temperature $T_z(z)$ under electrical and optical cooling, with fits to $\text{Eq.}\, (\ref{eq:T_vs_position})$, with $T_x$, $u_{\text{eq}}$, and $c_r$ replaced by $T_z$, $z_{\text{eq}}$, and $c_z$.}
	\label{fig:detection}
\end{figure}
We observe that under fixed electrical and optical feedback-cooling parameters, the steady-state temperature varies as a function of the particle's equilibrium position inside the Paul trap. To understand this dependence on position, we have to take into account two position-related effects: First, as described in Sec.~\ref{sec:position_optimization}, the signal of the particle's center-of-mass motion depends on the position of the particle. This position signal directly impacts the feedback signal.
Second, the electric and optical fields generated by the feedback actuators have their own spatial distributions, which depend on the feedback electrode geometry and on the \SI{1064}{\nano\meter} laser intensity profile, respectively. As a consequence, even for a fixed feedback signal, the feedback force depends on the particle position.

In $\text{Fig.}\, \ref{fig:detection}(\text{c})$, we plot $T_x$ for different lateral displacement $u_{\text{eq}}$ of the particle under both electrical and optical cooling at a pressure of \SI{2.1e-2}{\milli\bar} and a fixed feedback phase. The feedback gain was fixed to a sufficiently low value that the particle did not overheat in regions with negative damping.
For both electrical and optical feedback, the minimum temperature corresponded to the particle being positioned at the detection beam focus. At this point, the position signal is at its maximum, and with it, the feedback signal. Recall that the cooling and the detection lasers follow the same optical path, and thus the beams' foci also coincide, under the assumption that chromatic aberration of the focusing lens system can be neglected. Therefore, at the position for optimal detection, we also find the maximum of the radiation pressure force, and hence optimal optical cooling. Away from the focal point, cooling is less effective, as the particle is displaced to a region in which the detection response degrades. For both electrical and optical cooling, we also observe two maxima of $T_x$ above $T_0$, corresponding to heating in the region where the feedback signal experiences a phase shift of $\pi$ relative to the center of the beam, as described in more detail in Appendix~\ref{ap:conversion_factor}.

To model $T_x(u_{\text{eq}})$, we take into account an explicit position dependence of the feedback force (see Appendix \ref{ap:phase}).  We obtain
\begin{equation}
T_x(u_{\text{eq}})=\frac{\gamma_0T_0}{\gamma_0+g\ c_r(u_{\text{eq}})f_{\text{fb}}(u_{\text{eq}})},
\label{eq:T_vs_position}
\end{equation}
where $f_{\text{fb}}(u_{\text{eq}})$ is the spatial distribution of the feedback force and $g$ is the electronic gain set with the FPGA.
The position dependence of the detection amplitude and sign is expressed through the term $c_r(u_{\text{eq}})$.
The dependence of $f_{\text{fb}}(u_{\text{eq}})$ on $u_{\text{eq}}$ differs for electrical and optical forces: From numerical simulations of the feedback electrodes' fields, we find that the electrical feedback force $f_{\text{fb}}^{\text{el}}(u_{\text{eq}})$ varies only by $\sim 1 \%$ over the extent of the particle displacements explored here, and so we consider it to be constant. In contrast, the radiation pressure force exerted by the cooling beam is modeled as Gaussian: $f_{\text{fb}}^{\text{opt}}(u_{\text{eq}})\propto\exp(-2(u_{\text{eq}}-u_0)^2/s_0^2)$, where $s_0$ is the electric-field waist (1/$e$ radius) of the \SI{1064}{\nano\meter} laser.

We fit $\text{Eq.}\, (\ref{eq:T_vs_position})$ to $T_x(u_{\text{eq}})$ (solid lines in $\text{Fig.}\, \ref{fig:detection}(\text{c})$) with $g$, $u_0$ and $s_0$ as fit parameters. The model captures our experimental findings: The best cooling performance is achieved when the particle oscillates at the center of the detection beam. Here, the detection efficiency and the optical and electrical cooling strength are at their maximum. For particle displacements $|u_{\text{eq}}|>w_0$, the detection slope changes sign, and accordingly, the particle is heated by the feedback. For even larger displacements from the focus, the position and the feedback signals go to zero, and thus the particle approaches $T_0$. Under optical cooling, $T_0$ is reached earlier than with electrical cooling, due to the exponential decay of the radiation pressure cooling force.

We performed the same measurements for the axial mode: $\text{Fig.}\, \ref{fig:detection}(\text{d})$ shows $T_z(z_{\text{eq}})$ obtained with electrical and optical feedback cooling. For the case of optical cooling, the axial feedback force experienced by the particle derives from the beam's gradient force rather than from radiation pressure ($\text{Fig.}\, \ref{fig:setup}(\text{b})$). Accordingly, we fit $T_z(z_{\text{eq}})$ to $\text{Eq.}\, (\ref{eq:T_vs_position})$ with $f_{\text{fb}}^{\text{opt}}\propto(z_{\text{eq}}-z_0)\exp(-2(z_{\text{eq}}-z_0)^2/s_0^2)$, where $z_0$ is the axial focus position. The fit captures the sign flip of the axial gradient force when the particle is displaced across the focus, which has the consequence of switching $f_{\text{fb}}^{\text{opt}}$ from a cooling to a heating force. Optical axial cooling is optimized at a position displaced from the cooling beam focus. At this point, the amplitude of the particle's motion lies entirely in a region in which the direction of the optical gradient force remains constant.

The electrical cooling force, as for the radial modes, is independent of particle displacement, and thus the steady-state temperature as a function of position has the same characteristic shape as in $\text{Fig.}\, \ref{fig:detection}(\text{c})$.

\section{\label{sec:conclusion}Conclusions}
We have demonstrated 3D feedback cooling of a silica nanosphere levitated in a Paul trap down to temperatures of a few \si{\milli\kelvin} in a room-temperature experiment, which is three orders of magnitude lower than in previously reported experments~\cite{nagornykh2015, Conangla2018,delord2020} where cooling is also performed in the Paul trap potential. A cold damping technique was used, exploiting either electrical or optical forces as feedback actuators. We have characterized cooling performance for different background pressures in terms of the feedback force's amplitude and phase. Optical and electrical cooling yield similar results, with the temperature limit set by the back-action of the motion measurement and, for pressures of \SI{3.5e-4}{\milli\bar} and \SI{3e-5}{\milli\bar}, by background gas collisions.  For pressures of \SI{4.6e-7}{\milli\bar}, we see evidence for an additional heating mechanism.. Furthermore, a direct measurement of the response functions of our position detection system reveals the relationship between the particle's position and the cooling efficiency.

We expect the degree of control over experimental parameters developed here to be beneficial in various levitated-optomechanics applications. As one example, sub-wavelength cooling and position control in a Paul trap would allow one to localize the nanosphere inside an optical cavity, enabling several optomechanics protocols \cite{oriol2011,chang2010,carlos2019,henning2020, Millen_2020}.
Also, an optically cooled silica sphere could be used to sympathetically cool highly absorptive particles without the need for direct laser illumination, similar to the sympathetic cooling demonstrated for different ion species} \cite{Kielpinski}. The highly absorptive particles could be NV centers\cite{conangla2020_2} or nano-magnets.

In a new vacuum chamber assembled after these measurements, we use laser-induced acoustic desorption \cite{bykov2019} to load and trap nanoparticles at UHV pressures below $\SI{1e-10}{\milli\bar}$, an improvement of over three orders of magnitude with respect to the pressures discussed here. Together with the steps outlined in Sec.~\ref{sec:cooling:1axis:min}, namely, more efficient collection of scattered light and particle trapping at higher frequencies, we expect that it should be possible to enter the quantum regime.

\begin{acknowledgments}
	We thank Felix Tebbenjohanns, Martin Frimmer and Lukas Novotny for fruitful discussions. This work was supported by Austrian Science Fund (FWF) Project No. Y951-N36.
\end{acknowledgments}

\appendix
\section{Motion detection calibration}\label{ap:detection}
We model the nanosphere in the presence of feedback cooling as a set of three damped harmonic oscillators, one for each axis of the trap. We focus on the motion along $\hat{x}$, where
\begin{equation}
\ddot{x}+(\gamma_0+\gamma_{x})\dot{x}+\omega_x^2 x=\frac{F_{\text{th}}}{m}+\gamma_{x}\delta\dot{x}_{\text{il}}.
\label{eq:eom}
\end{equation}
Here $x$ represents the particle's displacement, $\omega_x$ is the angular frequency of oscillation, and $m$ is the mass of the particle. The interaction with background gas in the vacuum chamber is accounted for as a viscous force with damping rate $\gamma_0$ and as a Langevin force $F_{\text{th}}(t)$. Analogously, the feedback force is accounted for as a viscous force with damping rate $\gamma_{x}$ and as a Langevin force proportional to $\delta \dot{x}_{\text{il}}$.  In the main text, we refer to the damping rate of the feedback force as $\gamma_{\text{fb}}$, but here we explicitly consider the damping rates along each of the three axes, and so we introduce the new terms $\gamma_{x}$, $\gamma_{y}$, and $\gamma_{z}$.

From Eq.~(\ref{eq:eom}) one obtains $S_x$, the single-sided PSD of the particle's motion, in the absence of feedback cooling, that is, for $\gamma_{x}=0$:
\begin{equation}
S_x(\omega) =|\chi_x(\omega)|^2S_{F_{\text{th}}},
\label{eq:PSD_fb_off}
\end{equation}
where $\chi_x$ is the mechanical susceptibility of the damped harmonic oscillator along the $x$ axis,
\begin{equation}
\chi_x(\omega) = \frac{1}{m(\omega_x^2-\omega^2-2i\gamma_0\omega)},
\label{eq:chi_x}
\end{equation}
and $S_{F_{\text{th}}}$ is the PSD of $F_{\text{th}}$. Using the fluctuation-dissipation theorem, we obtain $S_{F_{\text{th}}}=4\gamma_0 mk_BT_0/\pi$. The analogous equations to Eq.~(\ref{eq:PSD_fb_off}) for $\hat{y}$ and $\hat{z}$ are
$S_y(\omega) =|\chi_y(\omega)|^2S_{F_{\text{th}}}$ and
$S_z(\omega) =|\chi_z(\omega)|^2S_{F_{\text{th}}}$,
where
\begin{align}
\chi_y(\omega) &= \frac{1}{m(\omega_y^2-\omega^2-2i\gamma_0\omega)}, \text{~and}
\label{eq:chi_y}
\\
\chi_z(\omega) &= \frac{1}{m(\omega_z^2-\omega^2-2i\gamma_0\omega)}.
\label{eq:chi_z}
\end{align}

We would now like to calibrate our detection for the particle's motion.
We obtain PSD data from a spectrum analyzer at a pressure of \SI{3.8e-2}{\milli\bar} and with no cooling.
In our experiment, the Paul trap's radial axes are oriented at $\theta\cong\SI{45}{\degree}$ with respect to the radial photodiode axis of detection, as depicted in Fig.~\ref{fig:setup}. The radial detector thus measures a signal $V_r(t)=c_r(\cos(\theta)x(t)+\sin(\theta)y(t))$, where $c_r$ is the conversion factor defined in Sec.~\ref{sec:position_optimization}. Accordingly, the PSD $S_r$ of the detected radial motion driven by $F_{\text{th}}$ is described by the double-peaked response function
\begin{equation}
S_{r(\text{det})}=(\cos^2(\theta)S_x+\sin^2(\theta)S_y)+S^n_{r(\text{ol})},
\end{equation}
where $S^n_{r(\text{ol})}$ is the single-sided PSD of the position-imprecision noise of the out-of-loop radial detector.
We fit the measured radial PSD with
\begin{equation}
S_{r(\text{fit})}(\omega)=(|\chi_x(\omega)|^2a_x+|\chi_y(\omega)|^2a_y)+r_{\text{noise}},
\end{equation}
with $a_x$, $a_y$, $\omega_x$, $\omega_y$, $\gamma_{0}$, and $r_{\text{noise}}$ as fit parameters, from which we obtain $\theta=\SI{47.39(1)}{\degree}$.
We then extract the absolute value of the radial calibration factor $|c_r|$ from the relation \cite{hebestreit2018} $S_{r(\text{fit})}=c_r^2S_{r(\text{det})}$.

In order to extract the axial calibration factor $c_z$, we note that the axial photodiode axis of detection is oriented parallel to $\hat{z}$. The PSD of the detected axial motion is described by $S_{z(\text{det})} = S_z+S_{z(\text{il})}^{n}$, where $S_{z(\text{il})}^{n}$ is the single-sided PSD of the position-imprecision noise of the in-loop axial detector. Note that we do not use a second out-of-loop detector for the $z$ axis. We fit the measured axial PSD with
\begin{equation}
S_{z(\text{fit})}=|\chi_z(\omega)|^2a_z + z_{\text{noise}},
\end{equation}
with $a_z$, $\omega_z$, $\gamma_{0}$ and $z_{\text{noise}}$ as fit parameters, and we extract the absolute value of the axial conversion factor $|c_z|$ from the relation $S_{z(\text{fit})}=c_z^2S_{z(\text{det})}$.

We now return to Eq.~(\ref{eq:eom}) and write down the PSD of the particle's motion in the presence of feedback cooling:
\begin{equation}
\begin{split}
	S_r^{\text{fb}}(\omega)=\cos^2(\theta)|\chi_x^{\text{fb}}|^2(S_{F_{\text{th}}}+m^2 \gamma_{x}^2 \omega^2 S^n_{r(\text{il})}/2)\\
+\sin^2(\theta)|\chi_y^{\text{fb}}|^2(S_{F_{\text{th}}}+m^2 \gamma_{y}^2 \omega^2 S^n_{r(\text{il})}/2)+S^n_{r(\text{ol})},
\end{split}
\label{eq:app_fb_on}
\end{equation}
where we define the modified susceptibilities $\chi_i^{\text{fb}}$, with $i=\{x,y,z\}$, by the substitutions $\gamma_0\to\gamma_0+\gamma_{i}$
in Eqs.~(\ref{eq:chi_x}), (\ref{eq:chi_y}), and (\ref{eq:chi_z}). 
The term $S^n_{r(\text{il})}$ is the PSD of the imprecision noise as measured with the radial in-loop detector. In practice, the two terms in Eq.\ (\ref{eq:app_fb_on}) proportional to $S^n_{r(\text{il})}$  can be considered as constants inside the bandwidth of the radial susceptibilities, since $\gamma_{x} \ll \omega_x$ and $\gamma_{y} \ll \omega_y$ within the range of the feedback gains used in our experiments. Therefore, we fit the measured radial PSD with
\begin{equation}
	S_{r(\text{fit})}^{\text{fb}}(\omega)=(|\chi_x^{\text{fb}}(\omega)|^2a_x^{\text{fb}}+|\chi_y^{\text{fb}}(\omega)|^2a_y^{\text{fb}})+b_{\text{noise}},
	\label{eq:rad_fit}
\end{equation}
with $a_x$, $a_y$, $\omega_x$, $\omega_y$, $\gamma_{x}$, $\gamma_{y}$ and $b_{\text{noise}}$ as fit parameters. For the case of axial feedback cooling, we fit the PSD of the measured signal with
\begin{equation}
S_{z(\text{fit})}^{\text{fb}}(\omega)=|\chi_z^{\text{fb}}(\omega)|^2a_z^{\text{fb}} + c_{\text{noise}},
\label{eq:ax_fit}
\end{equation}
with $a_z$, $\omega_z$, $\gamma_{z}$ and $c_{\text{noise}}$ as fit parameters. The radial and axial temperatures under feedback cooling are then calculated as
\begin{equation}
	\begin{split}
	T_x&=\frac{\gamma_0a_x^{\text{fb}}}{\gamma_xa_x}\times T_0,\\
		T_y&=\frac{\gamma_0a_y^{\text{fb}}}{\gamma_ya_y}\times T_0, \ \text{and}\\
		T_z&=\frac{\gamma_0a_z^{\text{fb}}}{\gamma_za_z}\times T_0.
	\end{split}
\end{equation}
In Paul traps, $T_0$ is not necessarily equal to the temperature of the buffer gas. A more detailed discussion about the value of $T_0$ can be found in Appendix~\ref{ap:calib_temp}.

\section{Calculating mass and charge from frequency jumps}\label{ap:jumps}
\begin{figure}[ht]
	\centering
	\includegraphics[width=1\linewidth]{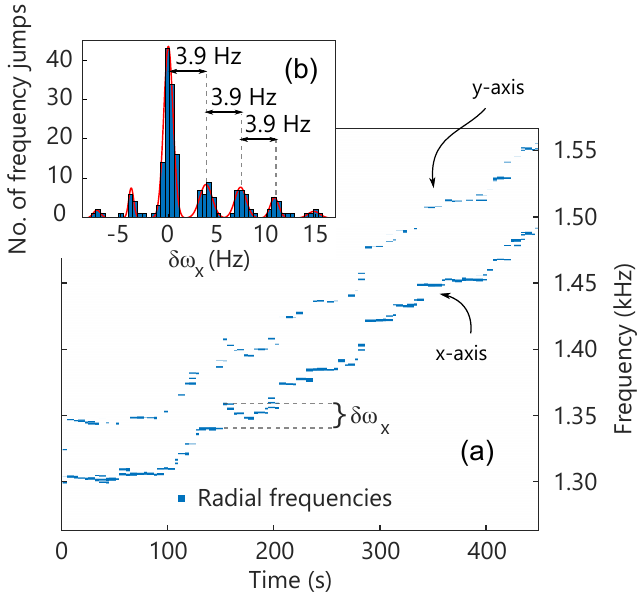}
	\caption{Variation in time of the radial secular frequencies of a trapped nanoparticle. The pressure gauge is turned on, charging the particle.  Negative values of $\delta\omega_x$ correspond to acquisition of negative charges by the particle. Inset: Histogram of frequency jumps with an amplitude $\delta\omega_x$. The red curve is a sum of Gaussian fits centered at the histogram's maxima.}
	\label{fig:jumps}
\end{figure}
The mass of a particle trapped in a Paul trap can be determined by varying the particle's charge and measuring its resonance frequencies ~\cite{arnold1979,schlemmer2001nondestructive}. After a nanoparticle has been trapped, we increase its charge using a Pirani/cold-cathode pressure gauge. The operating principle of the cold-cathode gauge is based on ionizing residual gas, and we hypothesize that the ionized gas leaks from the gauge volume and reaches the trapped particle~\cite{frimmer2017,bullier2020}. When the gauge is turned on, we observe discrete jumps in the motional frequencies of a trapped nanoparticle, indicating that the particle's charge state has shifted. In Paul traps, the resonant frequencies of oscillation $\omega_x$, $\omega_y$ and $\omega_z$ are related to the charge $Q$ on the trapped particle as~\cite{berkeland98}
\begin{align}
	&\omega_x \cong \frac{\Omega_\text{d}}{2} \sqrt{a_x + q_x^2/2}, \label{eq:omegax}\\
	&\omega_y \cong \frac{\Omega_\text{d}}{2} \sqrt{a_y + q_y^2/2}, \label{eq:omegay}\\
    &\omega_z \cong \frac{\Omega_\text{d}}{2}\sqrt{a_z}, \label{eq:omegaz}
\end{align}
where
\begin{align}
&a_x = -\frac{4Q}{m\Omega_\text{d}^2}\left(\kappa_{\text{end}} \frac{V_{\text{end}}}{z_0^2} + \kappa_{\text{RF}} \frac{V_{\text{off}}}{r_0^2} \right), \\
&a_y = -\frac{4Q}{m\Omega_\text{d}^2}\left(\kappa_{\text{end}} \frac{V_{\text{end}}}{z_0^2} - \kappa_{\text{RF}} \frac{V_{\text{off}}}{r_0^2} \right), \\
&a_z = \kappa_{\text{end}} \frac{8Q V_{\text{end}}}{m z_0^2 \Omega_\text{d}^2}, \label{eq:az}\\
&q_x = \kappa_{\text{RF}} \frac{QV_{\text{pp}}}{mr_0^2\Omega_\text{d}^2} \label{eq:qx} \\
&q_y = -q_x,
\end{align}
$\Omega_\text{d}$ is the driving frequency of the trap electrodes, $m$ is the mass of the particle, $\kappa_{\text{RF}}$ and $\kappa_{\text{end}}$ are geometric factors, $V_{\text{end}}$ is the DC voltage applied to the trap endcaps, $V_{\text{off}}$ is a DC offset voltage applied to the radial electrodes, $z_0$ is half the distance between the trap endcaps, $r_0$ is half the distance between opposing radial electrodes, and $V_{\text{pp}}$ is the peak-to-peak voltage applied to the trap electrodes. Our trap parameters are $\kappa_{\text{RF}} = 0.93$, $\kappa_{\text{end}} = 0.22$, $z_0 = \SI{1.7}{\milli\meter}$, and $r_0 = \SI{0.9}{\milli\meter}$.

In Fig.\ \ref{fig:jumps}, we plot the secular radial frequencies of a trapped particle measured over several hundred seconds, with the gauge kept on. We see a net increase of the frequencies, which we identify with positive ions from the gauge. We observe that as soon as the gauge is turned off, the frequencies remain constant. In our experimental setup, we select positively charged particles during the loading procedure by positively biasing the endcap electrodes. Subsequent charging with the pressure gauge is then used to boost the secular frequency.

In the inset of Fig.\ \ref{fig:jumps}, we plot the histogram of events that shift the particle frequency by $\omega_x\to\omega_x + \delta\omega_x$ as a function of the jump magnitude $\delta\omega_x$. The highest peak, at $\delta\omega=0$, indicates that for most of the measurement interval, the frequency remained at a constant value. All other peaks occur at integer multiples of a fundamental frequency $\delta\omega_x=\SI{3.9(6)}{\hertz}$. We interpret this discreteness in the magnitude of frequency jumps as direct evidence of the quantization of the charges gained or lost by the particle. We assume that the smallest frequency jump is produced by the gain of a single elementary charge $e$, which together with the total frequency shift $\Delta \omega_x$ allows us to calculated the number of elementary charges $N = \Delta \omega_x / \delta \omega_x$ obtained during the particle charging process. If the change in charge $\Delta Q = Ne$ is small compared to the initial charge $Q_0$, then $Q_0$ can be calculated as
\begin{align}
Q_0 = \frac{\omega_z\left(Q_0\right) \Delta Q }{2\Delta \omega_z}.
\label{eq:charge}
\end{align}
By substituting the total charge $Q_0$ and the trap parameters into Eqs.~(\ref{eq:az}) and (\ref{eq:omegaz}), the mass of the particle can be estimated. For example, for the particle used to obtain the data shown in Fig.~\ref{fig:tvs}, the mass is \SI{1.8(2)e-17}{\kilo\gram}.  This value is two-thirds of the mass specified by the particle manufacturer.

\section{Steady state temperature $T_0$ in the absence of feedback}\label{ap:calib_temp}
The harmonic motion of a particle in a Paul trap is only an approximation. In fact, the steady-state kinetic energy contains the energy of the secular motion and the energy of the micromotion. Moreover, even if we focus only on the secular motion, $T_0$ is not necessary equal to the temperature of the background gas~\cite{chen2014neutral} because the particle is subject to a time-dependent potential.

For $q\ll1$, the total kinetic energy is twice the energy of the background gas and is equally split between the secular motion and the micromotion. Therefore, $T_0 = \SI{300}{\kelvin}$ in this case. Due to technical limitations of our experimental setup, we had to work with relatively high $q$-parameters in order to obtain resonant frequencies on the order of few kilohertz. One reason to pursue these frequencies was that we needed strong enough confinement to keep the particle within the waist of the detection beam. Another reason was that we needed to spectrally separate motional modes. Therefore, in our experiments, the $q$-parameter reached values of $0.7$. Analytical expressions from Ref.~\cite{chen2014neutral} suggest that in this case, the total average kinetic energy, which includes the energy of the secular motion and the energy of the micromotion, is three times higher than the average kinetic energy of the background gas molecules and that $T_0$ has to be corrected by a factor of $1.3$. We have also performed more detailed numerical simulations of the motion of the particle trapped in a Paul trap and immersed in the background gas. The simulation gives us the same correction factor of 1.3. Note that this correction factor does not apply for motion along the $z$ axis since along this axis, the particle is confined with the electrostatic potential.

\section{Micromotion compensation}\label{ap:micromotion}
We compensate for excess micromotion~\cite{berkeland98} by aligning the average position of the particle in the radial plane to the RF null of the Paul trap with the help of the compensation electrodes. The measure of the quality of the compensation is the height of the peak at the drive frequency $\Omega_d$ in the spectrum of the particle motion. By careful positioning of the particle inside the trap, we bring the peak to the detection noise level, which means \SI{\sim 0.3}{\nano\meter} amplitude of the excess micromotion. In practice, however, the stray fields may change over time, and thus the particle may drift from the null point of the Paul trap, resulting in excess micromotion.

Excess micromotion can couple to the secular motion via collisions with the background gas. Therefore, we have studied numerically how sensitive the equilibrium secular CoM temperature is to the quality of the micromotion compensation. For this study, we displaced the particle from the trap center by \SI{4}{\micro\meter} (the range of displacements in Sec.~\ref{sec:position_vs_T}), calculated the secular temperature, and compared it with that of a particle exactly in the center of the trap. While the energy in the excess micromotion increased, we found no significant difference in the effective temperature of the secular motion.

\section{Optimal cooling phase and position dependent feedback forces}\label{ap:phase}
For low feedback gain, the temperature along the $x$ axis of motion is~\cite{Cohadon1999, li2011}
\begin{align}
	T_x&=\frac{m\omega_x^2}{k_B}\int_{0}^{\infty}S_x^{\text{fb}}d\omega \nonumber \\ &=\frac{\gamma_0T_0}{\gamma_0+\gamma_x^{\text{fb}(\phi,u_{\text{eq}})}},
	\label{eq:temperature_low_gain}
\end{align}
where $S_x^{\text{fb}}=\chi_x^{\text{fb}}S_{F_{\text{th}}}$ is the PSD of the particle's motion along $x$ under feedback cooling, $\chi_x^{\text{fb}}$ and $S_{F_{\text{th}}}$ have been defined in Appendix \ref{ap:detection}, and $\gamma^{\text{fb}}_x(\phi,u_{\text{eq}})$ is the feedback gain, which depends on the phase $\phi$ set with the FPGA and on the particle equilibrium position $u_{\text{eq}}$.
Equation (\ref{eq:T_vs_phase}) in the text is obtained by substituting $\gamma^{\text{fb}}_x(\phi)=-\gamma_{\text{fb}}\sin(\phi)$ into Eq.\ (\ref{eq:temperature_low_gain}). Equation (\ref{eq:T_vs_position}) is obtained by substituting $\gamma^{\text{fb}}_x(u_{\text{eq}})=gc_r(u_{\text{eq}})f^{\text{fb}}_x(u_{\text{eq}})$ into Eq.\ (\ref{eq:temperature_low_gain}).

\section{Conversion factors for different positions}\label{ap:conversion_factor}
\begin{figure}[ht]
	\centering
	\includegraphics[width=1\linewidth]{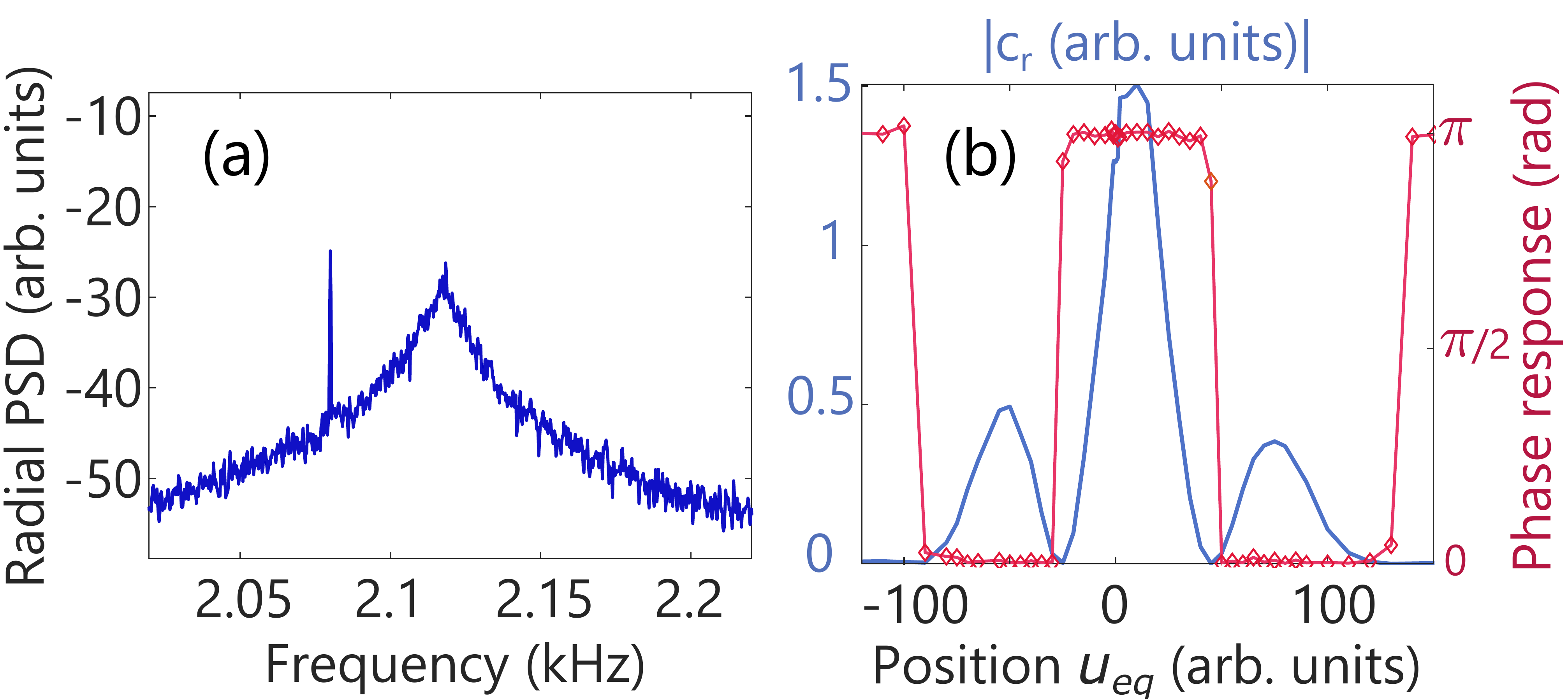}
	\caption{(a) Radial PSD showing the broad thermally driven resonance of the particle motion centered at $\omega_x$ and the narrow peaked response to the harmonic force drive close to resonance. (b) Blue: absolute value of the radial conversion factor $|c_r(u_{\text{eq}})|$ as a function of the lateral particle displacement $u_{\text{eq}}$. Red: phase difference between the harmonic force drive and the detected response of the particle at the drive frequency as a function of $u_{\text{eq}}$.}
	\label{fig:phase_responce}
\end{figure}
In Appendix \ref{ap:detection}, we have shown how to determine the absolute values of the conversion factors $|c_r|$ and $|c_z|$. The signs of the conversion factors are obtained by measuring the phase response of the particle motion to a coherent driving force: We drive the particle motion by supplying a sinusoidal electric signal through feedback electrodes. The frequency of this signal is tuned to one of the three trap resonances, chosen such that the particle responds to the force it experiences only along the corresponding axis of motion. An example PSD of the radial particle motion under the influence of the drive is shown in Fig.\ \ref{fig:phase_responce}(a). The plot in Fig.\ \ref{fig:phase_responce}(b) shows, in red, the phase difference between the radial drive signal and the detected particle's motion at the drive frequency as a function of the lateral displacement $u_{\text{eq}}$ relative to the detection beam. The blue line in the same plot is the absolute value of the radial calibration factor as measured for the same displacements $u_{\text{eq}}$. The plot was obtained by exciting the particle's motion along $x$, but similar results hold for the other two axes. We observe phase shifts of $\pi$ corresponding to the dips of $|c_r|$. These phase shifts do not originate from the particle motion, since the response to the drive signal does not depend on the particle position. Rather, we interpret a phase shift as a sign change of the calibration factor $c_r$. By knowing the conversion factor's sign and amplitude, we are able to reconstruct the detection response functions shown in Figs.\ \ref{fig:detection}(a) and (b), as explained in the text.

\bibliography{biblio_cooling}

\end{document}